\documentstyle[aps,prd,eqsecnum,epsf]{revtex}

\begin{document}

\title{A detailed study of quasinormal frequencies of the Kerr black hole\thanks{Pacs: 04.25.Nx, 04.30.-w, 04.30.Nk, 04.70.-s, 04.70.Bw}}
\author{Hisashi Onozawa\thanks{E-mail:onozawa@phys.titech.ac.jp},
\\
Department of Physics, \\
Tokyo Institute of Technology,\\
Oh-okayama, Meguro, Tokyo 152, Japan}

\date{October 18, 1996}

\maketitle

\begin{abstract}
We compute the quasinormal frequencies of the Kerr black hole
using a continued fraction method.
The continued fraction method first proposed by Leaver
is still the only known method stable and accurate for
the numerical 
determination of the Kerr quasinormal frequencies.
We numerically obtain not only the slowly
but also the rapidly damped
quasinormal frequencies and analyze
the peculiar behavior
of these frequencies at the Kerr limit.
We also calculate
the algebraically special frequency 
first identified by Chandrasekhar
and confirm that
it coincide with the $n=8$ quasinormal frequency only at
the Schwarzschild limit.
\end{abstract}

\section{Introduction}

The linear perturbations of the Schwarzschild spacetime
were first examined by Regge and Wheeler \cite{REG57},
who were trying to solve the stability problem of the black hole.
They derived what is now called the Regge-Wheeler equation,
a second order ordinary differential
equation similar to the Schr\"{o}dinger equation.
The perturbation radiates away at the speed of light satisfying
the wave equation and hence
it corresponds to the gravitational wave around
the black hole.
Therefore to examine the perturbations of the black hole
is equivalent to solving the scattering problem of
gravitational waves around the black hole.
Their pioneering work has led to the study of
the quasinormal modes because they dominates most processes
involving perturbed black holes.
Quasinormal modes were first found
by Vishveshwara \cite{VIS70b} and by Press \cite{PRE71}
through numerical computations of the time evolution of the
gravitational waves around the black hole.
External perturbations
excite the quasinormal modes which in turn appear
as damped vibration, emitting gravitational waves with
specific frequencies.
This has been shown in several other investigations
on the response of the black hole
to the external perturbations.
\cite{LEA86,SUN88,SUN90,AND95a}.
There have been several algorithms proposed to
determine these resonant frequencies of the black hole.
The first method presented by Chandrasekhar and Detweiler \cite{CHA75a}
was not sufficient to obtain accurate values of the quasinormal frequencies
because their scheme, namely, a direct numerical integration of the
Regge-Wheeler equation, was not stable,
especially for rapidly damped modes.
However, it is remarkable that they were able to obtain for the first time
values of the first few least damped quasinormal modes.

Meanwhile,
there have been some attempts to obtain the frequencies
analytically.
An inverse potential method by Mashhoon \cite{BOL84,FER84,LIU95}
was the first attempt to obtain the quasinormal frequencies
using a potential for which the spectrum is analytically
solvable. 
There is also a semianalytic approach, 
the WKB method used by Schutz and Will \cite{SCH85}, 
and later refined by Iyer and coworkers \cite{IYE87a,IYE87b}.
They have provided relatively easy and
intuitively understandable schemes for obtaining the
quasinormal frequencies.
The methods have also been applied to the Kerr \cite{IYE87c}
and Reissner-Nordstr\"{o}m black holes \cite{KOK88}
and have provided numerical values
for fundamental modes.
However, these algorithms 
cannot provide very accurate values for the frequencies
and break down completely for rapidly damped modes.

Leaver \cite{LEA85} has shown that a continued fraction method
previously used for determining the energy spectrum of the
hydrogen molecule ion \cite{JAF34,BAB35} can be generalized to
determine the quasinormal frequencies.
This method provides extremely accurate values for
the frequencies of the
Schwarzschild and Kerr black hole.
It was found in the Schwarzschild case
that there are an infinite number of damped modes for every multipole
index $l$ and these modes
are located approximately parallel to the imaginary axis.
Also for the Kerr black hole, he calculated how these modes
move from the Schwarzschild quasinormal frequencies as the
black hole parameter changes.
Later, he generalized his method
to the Reissner-Nordstr\"{o}m
black hole \cite{LEA90}, and found out how the quasinormal
modes move as the charge of the black hole changes.
These results for the non-rotating cases
are confirmed by other independent investigations:
Nollert and Schmidt \cite{NOL92}
presented the Laplace transform method
which are in excellent agreement with Leaver's results for
the Schwarzschild black hole.
Andersson and coworkers 
used the phase integral method 
for the Schwarzschild black hole \cite{AND92a,And92b,AND93c}
to obtain the consistent results.
The last method was also used for the Reissner-Nordstr\"{o}m black
hole \cite{AND93b,AND94}.

There are several methods to precisely
obtain the quasinormal frequencies for the non-rotating cases, nevertheless, 
the continued fraction method by Leaver is still the only known
way which can be generalized to the Kerr quasinormal frequencies.
The Kerr black hole is most important astrophysically,
thus its quasinormal modes should be investigated in detail.
At the same time, 
several questions were raised by the previous detailed studies
of the quasinormal frequencies of the Reissner-Nordstr\"om
black hole \cite{ONO96,ONO96c,ONO96b}.
In Ref. \cite{ONO96} we have improved
a continued fraction method for the extreme case and then
discovered
a curious fact that the gravitational
and the electromagnetic frequencies are identical.
Whether this is also the case for the Kerr black hole is still unknown.
In Ref. \cite{ONO96c}
we presented a detailed calculation of the quasinormal modes
of charged black holes focusing on the
rapidly damped modes for nearly extreme cases.
We found that the rapidly damped modes show several peculiar
features:
the higher modes generally spiral into the value for the
extreme black hole as the charge increases.
We also found that there is a quasinormal mode
that converges to the algebraically special mode \cite{cspec}
when the charge vanishes, even though
there was no correlation between these two modes
in the charged case.
It is still unresolved whether this is just a coincidence
or can be explained physically.
There are arguments for and against the existence of
the quasinormal modes on the imaginary axis.
Some methods indicate that there should be such a mode and
that it corresponds to the ninth quasinormal mode.
The problem is that the algebraically special mode of
non-rotating black holes is purely imaginary and that 
most existing methods do not work for
purely imaginary modes.
In order to resolve this arguments, it will be useful
to investigate the Kerr case because the
algebraically special mode move away from the
imaginary axis
and thus can be compared with the quasinormal modes.
It is also important to know in detail the behavior of
quasinormal modes for nearly extreme Kerr black holes,
because
some astrophysically relevant black holes are a consequence 
of the coalescence of binary neutron stars
and are believed to have a considerably large angular momentum.
However, the behavior of these modes near the extreme limit is not
yet well known.

In this paper, the quasinormal frequencies of the Kerr black hole
are numerically computed using Leaver's continued fraction method
beyond his original calculation.
Several unresolved problems about the Kerr quasinormal modes
are discussed.

\section{Continued Fraction Equations}
\label{sec:ee}

It is well known that
perturbations of the Kerr black hole are described by the Teukolsky
equations.
In case a $t$- and $\varphi$- dependence is given by 
$e^{-i\omega t +i m \varphi}$, 
the separated differential equation for an angular part of perturbations is
\begin{equation}
  (1-u^2) \frac{d^2S_{s}}{du^2} - 2u\frac{dS_{s}}{du} + W(u)S_{s}=0,
\end{equation}
\begin{equation}
  W =  a^2 \omega^2 u^2 - 2 a \omega s u +
    A_{lm} + s  - \frac{(m+su)^2}{1-u^2},
  \label{eq:teu-ang}
\end{equation}
and that of a radial part is 
\begin{equation}
  \Delta \frac{d^2R_{s}}{dr^2} + 2(s+1)(r-M)\frac{dR_{s}}{dr}
  +V(r) R_{s} =0,   \label{eq:teu-rad}
\end{equation}
\begin{equation}
   V = \frac{K^2-2isK(r-M)}{\Delta} + 4 i s \omega r 
       -A_{lm} +2a\omega m -a^2\omega^2,
\end{equation}
where 
\begin{eqnarray}
  u &=& \cos \theta ,\\
  \Delta &=& r^2-2Mr+a^2, \\
  K &=& (r^2+a^2)\omega - am.
\end{eqnarray}
These equations correspond to the electromagnetic waves in the Kerr
geometry when $s=-1$ and the gravitational waves when $s=-2$.
In what follows, we normalize $2M=1$ if otherwise mentioned.

In the Schwarzschild case, Eq.(\ref{eq:teu-ang}) can be solved
analytically and hence the separation constant $A_{lm}$ can be
obtained as $A_{lm}=l(l+1)-s(s+1)$.
However, in the Kerr case, we have to solve it
numerically as follows \cite{LEA85}.
Boundary conditions for Eq.(\ref{eq:teu-ang}) are that $S_{s}$ is
regular at the regular singular points $u=\pm1$.
The indices there are determined by the regularity at both singular
points and are given by $\pm(m+s)/2$ at $u=1$ and $\pm(m-s)/2$ at $u=-1$.
A solution to Eq.(\ref{eq:teu-ang}) can be expressed as
\begin{equation}
  S_{s}(u)= e^{-a\sigma u} (1+u)^{|m-s|/2} (1-u)^{|m+s|/2}
               \sum_{n=0}^{\infty} a_n (1+u)^n.
\end{equation}
The expansion coefficients are related by a three-term recurrence
relation
and the boundary condition at $u=1$ is satisfied only by
its minimal solution sequence.
The recurrence relation
is
\begin{eqnarray}
&&  \alpha_0 a_1 + \beta_0 a_0 =0, \\
&&  \alpha_n a_{n+1} + \beta_n a_n + \gamma_n a_{n-1} =0, \hspace{5mm}(n \ge 1),
\end{eqnarray}
where the recurrence coefficients are, with $k_1=|m-s|/2$ and
$k_2=|m+s|/2$,
\begin{eqnarray}
  \hat{\alpha}_n &=& -2 (n+1)(n+2k_1+1),\\
  \hat{\beta}_n &=& n(n-1)+2n(k_1+k_2+1+2a\sigma) \nonumber \\
      & & +2a\sigma (2k_1+s+1)+(k_1+k_2)(k_1+k_2+1) \\
      & & -a^2 \sigma^2 - s(s+1)-A_{lm}, \\
  \hat{\gamma}_n &=& 2 a\omega ( n+k_1+k_2+s).
\end{eqnarray}
The minimal solution sequence will be obtained if the angular
separation constant $A_{lm}$ is a root of the continue fraction
equation below,
\begin{equation}
  0=\hat{\beta}_0-\frac{\hat{\alpha}_0\hat{\gamma}_1}{\hat{\beta}_1-}
            \frac{\hat{\alpha}_1\hat{\gamma}_2}{\hat{\beta}_2-}
            \frac{\hat{\alpha}_2\hat{\gamma}_3}{\hat{\beta}_3-}
            \frac{\hat{\alpha}_3\hat{\gamma}_4}{\hat{\beta}_4-}\cdots
   \label{eq:cf-r}
\end{equation}

The boundary conditions for quasinormal modes for the radial equation are 
\begin{eqnarray}
  R_{-2} &\sim& (r-r_{+})^{-s-i\sigma_{+}} e^{i\omega r},
   \hspace{5mm} \mbox{as} \hspace{5mm} r \rightarrow r_{+}, \\
  R_{-2} &\sim& r^{-1-2s+i\omega} e^{i\omega r}, 
   \hspace{5mm} \mbox{as} \hspace{5mm} r \rightarrow \infty,
\end{eqnarray}
where $\sigma_+  = (\omega r_+ -am) / b$  and $b=\sqrt{1-4 a^2}$.
Thus, the solution should be 
\begin{equation}
  R_{-2} = e^{i\omega r} (r-r_{-})^{-1-s+i\omega + i\sigma_{+}}
     (r-r_{+})^{-s-s\sigma_{+}}
       \sum^\infty_{n=0} a_{n}\left( \frac{r-r_+}{r-r_-} \right)^n,
         \label{eq:serk}
\end{equation}
where the expansion coefficients are again defined by a
three-term recurrence relation:
\begin{eqnarray}
&&  \alpha_0 a_1 + \beta_0 a_0 =0, \\
&&  \alpha_n a_{n+1} + \beta_n a_n + \gamma_n a_{n-1} =0,
      \hspace{5mm}(n \ge 1).
\end{eqnarray}
The recursion coefficients are
\begin{eqnarray}
  \alpha_n &=& n^2+(c_0+1)n+c_0, \\
  \beta_n  &=& -2n^2+(c_1+2)n+c_3, \\
  \gamma_n &=& n^2+(c_2-3)n+c_4-c_2+2,
\end{eqnarray}
and $c_n$ are defined by
\begin{eqnarray}
  c_0 &=& 1-s-i\omega-\frac{2i}{b}\left(\frac{\omega}{2}-am \right), \\
  c_1 &=& -4+2i\omega(2+b)+\frac{4i}{b}\left(\frac{\omega}{2}-am \right),\\
  c_2 &=& s+3-3i\omega-\frac{2i}{b}\left(\frac{\omega}{2}-am \right), \\
  c_3 &=& \omega^2(4+2b-a^2)-2am\omega-s-1+(2+b)i\omega \nonumber \\
      && -A_{lm}+
         \frac{4\omega+2i}{b}\left( \frac{\omega}{2}-am \right), \\
  c_4 &=& s+1-2\omega^2-(2s+3)i\omega-\frac{4\omega+2i}{b}
            \left( \frac{\omega}{2}-am \right).
\end{eqnarray}
The boundary condition that Eq.(\ref{eq:serk}) converges
as $r\rightarrow\infty$ reduces again to
\begin{equation}
  0=\beta_0-\frac{\alpha_0\gamma_1}{\beta_1-}
            \frac{\alpha_1\gamma_2}{\beta_2-}
            \frac{\alpha_2\gamma_3}{\beta_3-}
            \frac{\alpha_3\gamma_4}{\beta_4-}\cdots
   \label{eq:cf-a}
\end{equation}
The problem is now to seek frequencies which
satisfy both Eq.(\ref{eq:cf-r}) and Eq.(\ref{eq:cf-a}).

\section{Numerical Results}

\subsection{The case of $m=0$}

The case of $m=0$ is similar to the Reissner-Nordstr\"om
quasinormal modes
and hence several things will be repeated.
First, each $m=0$ mode always has a corresponding mode
symmetric about the imaginary axis.
This is due to a symmetry of Eqs. 
(\ref{eq:cf-r}) and (\ref{eq:cf-a}).
It is obvious that these continued fraction equations are
equivalent even after the transformation of
$m\rightarrow -m$, $i\omega \rightarrow (i\omega)^*$,
and $A_{lm} \rightarrow A_{lm}^*$.
Therefore, when $m=0$ a solution always appear as
a complex conjugate pair in $(i\omega)$.
In contrast to the Reissner-Nordstr\"om black hole
the mode-frequencies of the first few modes
have the damping rates (imaginary parts of $\omega$) which 
decrease monotonically as $a/M$ increases.
On the other hand, the oscillation frequencies
(real parts of $\omega$) increase with $a/M$,
which is the same as the Reissner-Nordstr\"om black hole.
These results are clear from Fig. 1, where we show the behavior of the
first three modes for $(-s,l,m)=(1,1,0)$, $(2,2,0)$, $(1,2,0)$,
and $(2,3,0)$. The left endpoint of each line corresponds to the
Schwarzschild mode, and the right endpoint corresponds to the $a/M=0.99$
Kerr mode.
The electromagnetic modes of the Reissner-Nordstr\"om,
which correspond to the $s=-1$ modes here,
moved very fast when the charge of the black hole is changed \cite{ONO96c}.
However the $s=-1$ Kerr black hole modes
move slowly even when the angular momentum of the
black hole is changed.
This can be understood,
because the electromagnetic perturbation depends
more on the change of the black hole electronic charge
rather than the change of the angular momentum.
Even though one might expect that the gravitational and the electromagnetic
mode coalesce in the extreme limit as was established in the
charged black hole case \cite{ONO96,ONO96b},
obviously there is no such correlation
between these two modes in Fig. 1.

The behavior for the first few modes were rather simple, but
the rapidly damped modes show the same strange behavior as in the
Reissner-Nordstr\"om black hole.
As presented in Fig. 2,
the first strange feature appears for $n=3$ in the $l=1$ electromagnetic modes
and for $n=6$ in the $l=2$ gravitational modes.
These modes go through a couple of small loops when the angular momentum
increases.
Similar loops are also found in the results for the higher mode number, $n$.
In these figures it is clear that 
the loops are getting bigger and the number of cycles increases
when the mode number is getting higher,
as in the case of the Reissner-Nordstr\"om black hole.
These strange features typically
appear when the ratio of the damping rate
to the oscillation frequency reaches a certain value, approximately
between five and ten for any multipole index $l$,
which is rather large compared with the charged black hole case.
Since the highly damped modes are not well understood, it is difficult
to explain why such a peculiar behavior occurs.
In the previous paper \cite{ONO96c} we suggested that the fall-off rate of the
effective potential at the horizon might be the cause.
This explanation suggests itself also in this case,
because for the Kerr black hole
the fall-off rate of the potential depends on the Kerr parameter
in a similar way as the Reissner-Nordstr\"om black hole.

It should be noted that we have tracked the trajectory of 
the $l=2$ ninth overtone ($n=8$) gravitational quasinormal mode in Fig. 2.
This frequency is known to coalesce with the algebraically special mode
at the Schwarzschild limit which is located on the imaginary axis.
The algebraically special mode is defined in a completely different
way from the quasinormal mode.
It is defined as a mode that is purely ingoing at both boundaries.
Thus, it seems that there is no reason that these modes have any
kind of relationship.
We will discuss this matter in Sec. \ref{sec:alg}.

The numerical error in the previous work \cite{ONO96c},
possibly caused by the bad convergence property
of the continued fraction,
does not effect the Kerr black hole case.
The convergence of the continued fraction is very fast for the
Kerr black hole and thus can avoid numerical errors.
When checking our code for the fundamental $l=2$ gravitational mode,
it was possible to obtain the mode accurately until $a/M=0.999995$.
For the Reissner-Nordstr\"om black hole the same mode could be
calculated only up to $Q/M=0.9985$.
Thus, it is concluded that although the
Teukolsky equations are more complicated than the Zerilli-Moncrief
equations the quasinormal frequencies can be calculated
very precisely through the continued fraction method even when the
Kerr parameter is near the maximum value.
The recurrence relations in the Reissner-Nordstr\"om black hole
are four-term recurrence relations and
numerical instability might be caused through
the transformation of four-term relations into three-term relations.

\subsection{The cases of $m \neq 0$}

First we discuss the case of the $s=-2$ gravitational perturbation
with the least multipole index $l=2$. ( The case with
higher multipole index is similar. )
It is noted that $m$ must be in a range of $-l \leq m \leq l$,
thus we have four different modes for $l=2$ with $m \neq0$.
Owing to the symmetry that we mentioned before,
all the modes with the azimuthal multipole index $m$ are obtained
by reflecting the modes with the index $-m$ about the imaginary axis.
We therefore discuss only the modes in the right-half of the
complex plane.
The trajectories of the $m=+1$ modes are plotted in Fig. 3.
We can find that the oscillation frequency becomes very fast
when the Kerr parameter is increased, 
on the other hand, the damping rates become slower.
The mode moves fast away from the corresponding Schwarzschild mode.
However the branch of $m=-1$ moves slowly and stays
near the corresponding Schwarzschild modes.
It is surprising that the $m=-1$ mode turns sharply 
at some specific Kerr parameter approximately $a=0.2$,
which is not yet found for any non-rotating black hole.
It is also noted that the mode of $m =\pm 1$ 
does not experience loops unlike a highly damped mode of the $m=0$ mode.
This distinguishes the behavior of $m \neq 0$ modes from that of
the $m=0$ modes.

The behaviour of the $m=\pm 2$ modes is quite similar to that of
the $m = \pm 1$ modes.
Fig. 4 shows how these modes move as the Kerr parameter increases.
As is easily seen, the $m=+2$ modes move very fast compared with any
other modes and these modes converge to a specific
undamped frequency at the Kerr limit.
It is already known that the modes of the $l=2, m=+2$ branch
accumulate onto the critical frequency of
super-radiance, $\omega_c=2$.
Detweiler has shown that when $a=1/2$
an infinite number of modes cluster at the critical
frequency.
This suggests a marginally
unstable feature of the extreme Kerr black hole
discussed by several authors \cite{DET80,DET83,SAS90},
and this is quite interesting to investigate in detail.
At present people believe that when the
Kerr parameter is close to the maximum
the Kerr parameter is decreased emitting the gravitational
waves and losing rotational energy.
Then the black hole settles down to
a stable non-extreme Kerr black hole.
You can also notice in Fig. 4 that
not all the modes converge to the real critical frequency,
but the sixth mode turns sharply at about $a=0.365$ and then limits to
a complex value around $\omega = (0.239, -0.242)$.
Detweiler's result was that there are an infinite number of
quasinormal modes converging to the critical frequency.
Our numerical results suggest that 
most of modes converge into the critical frequency
but some modes still stay away from the critical frequency.
We so far have calculated the modes up to $n=20$ but there was no other
mode which does not converge into the critical frequency.

It is also interesting that 
there seems to be some relation between the $m = \pm2$ modes and 
the $m = \pm1$ modes.
As is shown in Fig. 5,
the both $m=+1$ and $m=+2$ modes  start from the
Schwarzschild mode initially in the same direction and
the both $m=-1$ and $m=-2$ modes moves in the opposite direction to the
above two modes.
Therefore the trajectories of these two modes are tangent to 
each other.
On the other hand, the $m=0$ mode moves into 
a non-correlated direction.
It seems that the $m=\pm 2$ modes move just twice as fast as the $m=\pm1$ mode
initially.
(This can be also confirmed by examining the $l=3$ case, where the $m=3$
mode moves three times as fast as the $m=1$ mode.)
These feature can be understood,
the $m=\pm 2$ modes are more distorted than the $m=\pm 1$ modes and hence
the $m=\pm 2$ are more sensitive to the change of the Kerr parameter
that corresponds how much the spacetime is twisted.

Fig. 6 is a figure of the trajectories of the $l=1, m=\pm 1$ modes for the
electromagnetic perturbations ($s=-1$).
It is very similar to the case of the $-s=m=2$ modes.
These modes also converges into the undamped critical frequency
$\omega_c=1$.
Owing to Detweiler's argument,
for any $s$ the $l=m$ modes
have a similar accumulation of an infinite number of the quasinormal modes
onto the critical frequency $\omega=l$.
There was one mode which never converge into the critical frequency
for the $s=-2$ case, however for the $s=-1$ case
no such mode was found up to $n=8$.

\subsection{Algebraically special modes}
\label{sec:alg}

Here we only consider the algebraically special mode of the $s=-2$
gravitational perturbations, because the only known implication that
there is a correlation between
the algebraically special mode and the quasinormal mode
is that both modes for the Schwarzschild black hole
coalesce for gravitational perturbations.

The algebraically special perturbations were first considered
by Wald \cite{WAL73}, who
was interested in how the perturbation
can be the only independent component among several Weyl scalars.
(Actually it is impossible if you consider only the real frequency.)
As is well known,
the perturbations have generally two independent components $\Psi_0$ and
$\Psi_4$, which correspond to the ingoing and the outgoing wave, respectively.
Thus the algebraically special mode is the mode that has only an
ingoing or outgoing wave.
Such kind of perturbations were first
analytically solved by Chandrasekhar \cite{cspec}.
For the Schwarzschild black hole,
algebraically special frequency can be easily solved
by imposing the condition that Starobinsky's constant vanishes. (Because
two modes are related to each other with a Starobinsky constant.)
It is then found that there is such a mode on the imaginary axis
at $\omega=- 2 i / M$.

For the Kerr black hole it is rather difficult to obtain the
algebraically special frequencies
owing to the difficulty to get the separation constant $A_{lm}$
without resorting to numerical computation.
Nevertheless, if we use the continued fraction technique for the angular
part of the Teukolsky equations, the separation constant is accurately
obtained.
The square of the Starobinsky constant is given by
\begin{eqnarray}
 |C|^2 &=& \lambda^2(\lambda+2)^2
      -8\omega^2\lambda(\alpha^2(5\lambda+6)-12a^2)
 +144\omega^2(M^2+\sigma^2+\alpha^4),
 \label{eq:staro-c}
\end{eqnarray}
where
\begin{eqnarray}
 \alpha &=& a^2-\frac{am}{\omega},
\end{eqnarray}
and $\lambda$, the separation constant in Chandrasekhar's notation,
is related to ours with the relation
\begin{equation}
  \lambda=A_{lm}-2a\omega m+a^2\omega^2.
\end{equation}
Therefore we only need to solve $|C|^2=0$ together with Eq. (\ref{eq:cf-a}).
In contrast to quasinormal modes,
algebraically special modes can be calculated even when the black
hole is extreme, because we do not need to solve
Eq. (\ref{eq:teu-rad}) which has two confluent singularities
in the extremal limit.

The solutions always appear as a pair, owing to a
reflection symmetry about the real axis.
It is also noticed that the symmetry of Eqs. (\ref{eq:staro-c}) and
(\ref{eq:cf-a}) allows us to get the $m=-1, -2$ modes by reflecting 
the $m=+1, +2$ modes about the imaginary axis as in the quasinormal mode.
Therefore, it is enough to see the first quadrant.
The solutions of $l=2$ branch
which are obtained using the continued fraction scheme are plotted
in Fig. 7 together with Chandrasekhar's result \cite{cspec},
and these values are listed in Table 1.
The $m=0$ mode increases along the imaginary axis
initially as the Kerr parameter increases and then 
goes away from the imaginary axis when $a=0.247222$.
In Ref. \cite{cspec}
the value of the $m=0$ mode for $a=0.25$ still stays
on the imaginary axis, but Chandrasekhar mentioned that
this mode does not satisfy a positivity of $\kappa$.
This contradiction was caused, because his values were obtained using a
separation constant with large numerical error, and thus his value was wrong.
Our values are all in agreement with this positivity check,
since our scheme can avoid numerical errors.
If you see Fig. 6, there is no $m=-1$ and $m=-2$ mode 
in the first quadrant, and 
the $m=+1$ and $m=+2$ modes 
go horizontally from the imaginary axis
like the quasinormal mode.
Despite that, there seems to be 
no correlation with any quasinormal mode.

From the point of view of the scattering problem,
a quasinormal mode is regarded as a
pole of the reflection amplitude $R(\omega)$ and
the transmission amplitude $T(\omega)$, 
on the other hand, an algebraically special mode
corresponds to a zero of $R(\omega)$.
Therefore these two modes will never coalesce in general, which
is consistent with the above numerical results.
Superficially it might appear that these two modes coalesce
at the Schwarzschild limit,
but we believe that the quasinormal mode disappears at the limit
owing to a cancellation with the corresponding mode
in the left-half of the complex plane.

\section{Concluding Remarks}

In this paper we have presented several hints to be useful for
resolving a couple of questions concerning the
quasinormal mode of the black hole.
We have presented the results of detailed 
calculations of the quasinormal modes of the Kerr black hole.
This supplements previous studies in several ways.
First, we take the rapidly damped mode into consideration.
Second, we also investigate precisely how these modes
move as the Kerr parameter approaches the limit.
Third, we also consider the relationship between the
algebraically special mode and the quasinormal mode.

As was the case for the Reissner-Nordstr\"om black hole \cite{ONO96c},
the highly damped $m=0$ Kerr quasinormal modes
generally spiral into the extreme values.
Though not presented in this paper,
we have found the similar behavior
also in the case of the test scalar field
on the Kerr background ($s=0$).
Therefore these spirals
must be one of the fundamental features for rapidly damped modes.
The behavior of the $m \neq 0$ modes are, on the other hand, different from
the behavior of the $m=0$ mode.
The damping rates of $0 < m \leq l$ modes decrease very fast without
going through spirals.
An infinite number of the $m=l$ modes tend to accumulate onto
the critical frequency, $\omega_c=l$, as $a/M \rightarrow 1$.
We also found that there is one mode which does not converge
to the frequency.
Whether another mode like that exists is an open question.
At present there is no evidence up to $n=20$ for $l=2$ gravitational
perturbations.
The $-l \le m < 0$ modes have rather simple behavior in contrast to
the positive $m$ cases.
These modes move slowly around the corresponding Schwarzschild modes.
A interesting fact is that some of the modes go through a quick turn.
These surprising features for $m \neq 0$
were not yet found for the spherical
black holes and hence should be considered in detail.
Since the $m \neq 0$ modes do not exist for non-rotating black holes
(all the $m \neq 0$ modes degenerate into the $m=0$ case
owing to the spherical symmetry),
these modes are actually specific for the rotating black holes.
We also obtained the precise values of 
the algebraically special modes using a continued fraction method.
There was no
correlation between the algebraically special mode and
the quasinormal mode except that these modes
superficially coalesce in the Schwarzschild limit.
Currently we do not know why the coalescence occurs.
However, owing to the existence of the algebraically special mode 
at $\omega=-4i$ the quasinormal mode can not be there.
Two quasinormal modes coalesce and then probably cancel at the limit.
This cancellation prevents the existence of the purely
damped quasinormal mode.

\section*{Acknowledgements}

I would like to thank Professor A.~Hosoya for his continuous encouragement.
I wish to thank Professor H.~Ishihara for stimulating discussions.
This work is supported in part by
the JSPS Research Fellowships for Young Scientists and
the Scientific Research Fund of the Ministry of Education.


\newpage

\begin{figure}
\begin{center}
  \leavevmode
  \epsfysize=10.0cm
  \epsfbox{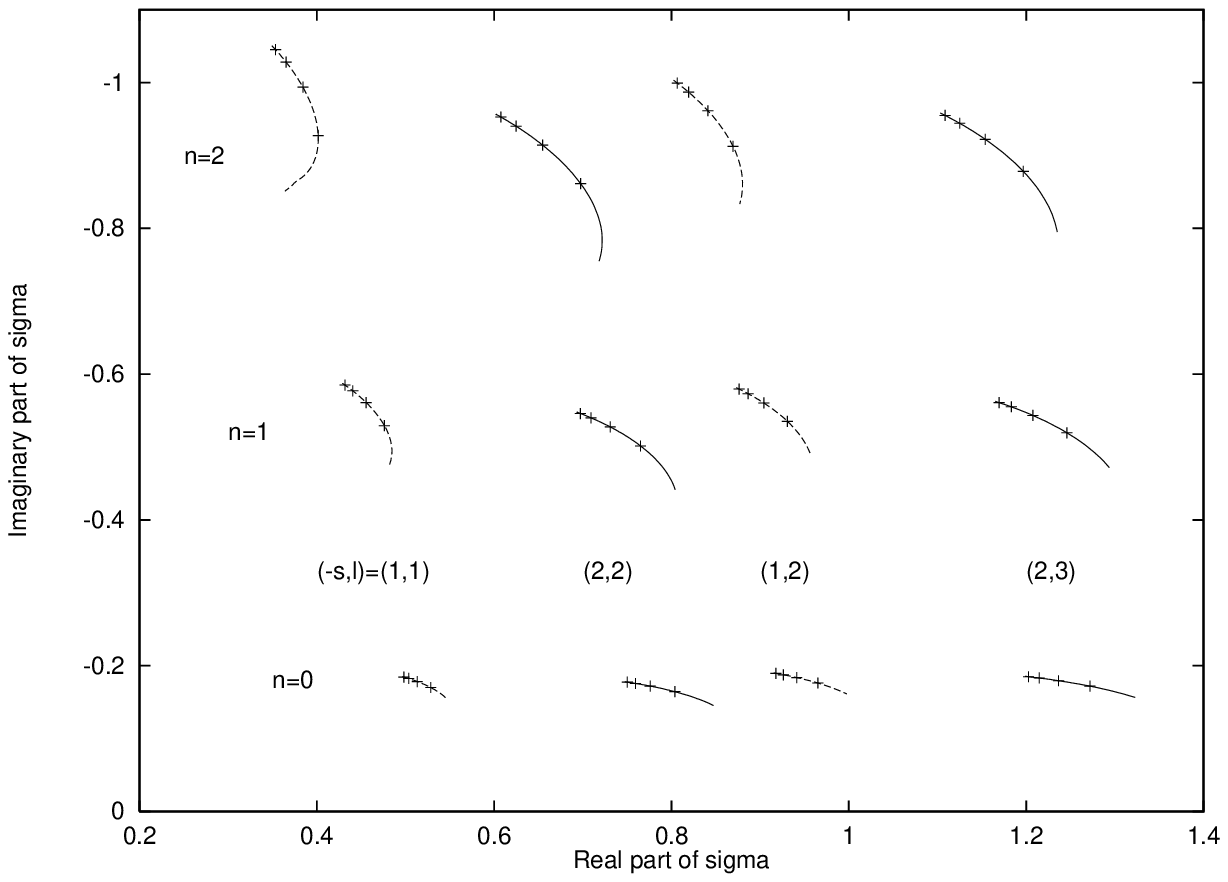}
\end{center}
\caption{
Solid lines and dashed lines are trajectories of the m=0
quasinormal frequencies of the gravitational and electromagnetic wave,
respectively.
Each right endpoint corresponds to the Schwarzschild quasinormal frequency,
and each right endpoint corresponds to
the frequency in the limit of nearly maximal angular momentum a/M=0.99.
Ticks marked on each line
are the frequencies for a=0.1, 0.2, 0.3 and 0.4 from the left to right.
}
\end{figure}

\newpage

\begin{figure}
\begin{center}
  \leavevmode
  \epsfysize=16.0cm
  \epsfbox{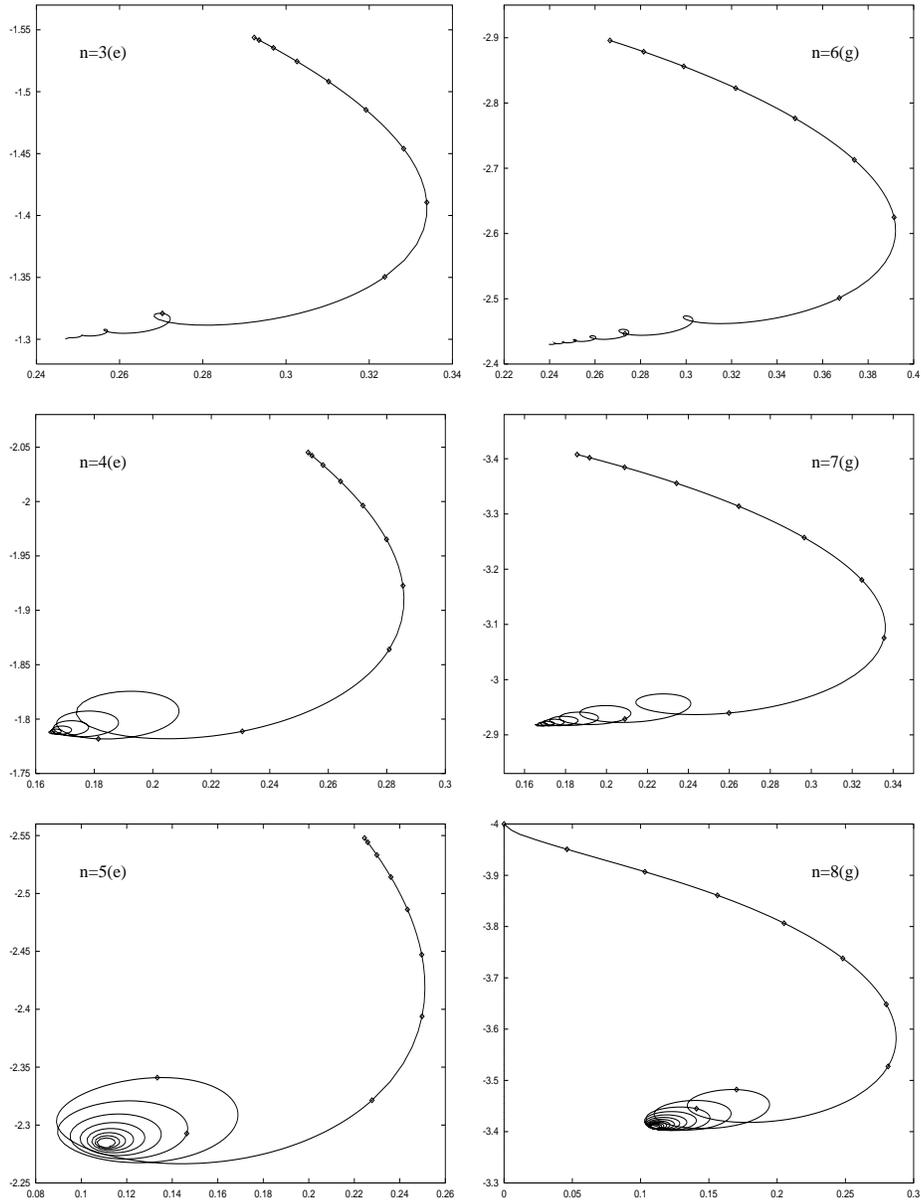}
\end{center}
\caption{
The first three specific quasinormal modes for m=0 are plotted.
The marks correspond to the frequencies of a=0, 0.05, 0.10, ..., 0.45.
These modes are generally spiral into the values in the extreme limit.
The qualitative features for the gravitational 
and electromagnetic frequencies are basically identical.
These figures are much finer than the figures in the previous
work for the charged black hole
because the continued fraction scheme works fine with
less numerical errors.
}
\end{figure}

\newpage

\begin{figure}
\begin{center}
  \leavevmode
  \epsfysize=10.0cm
  \epsfbox{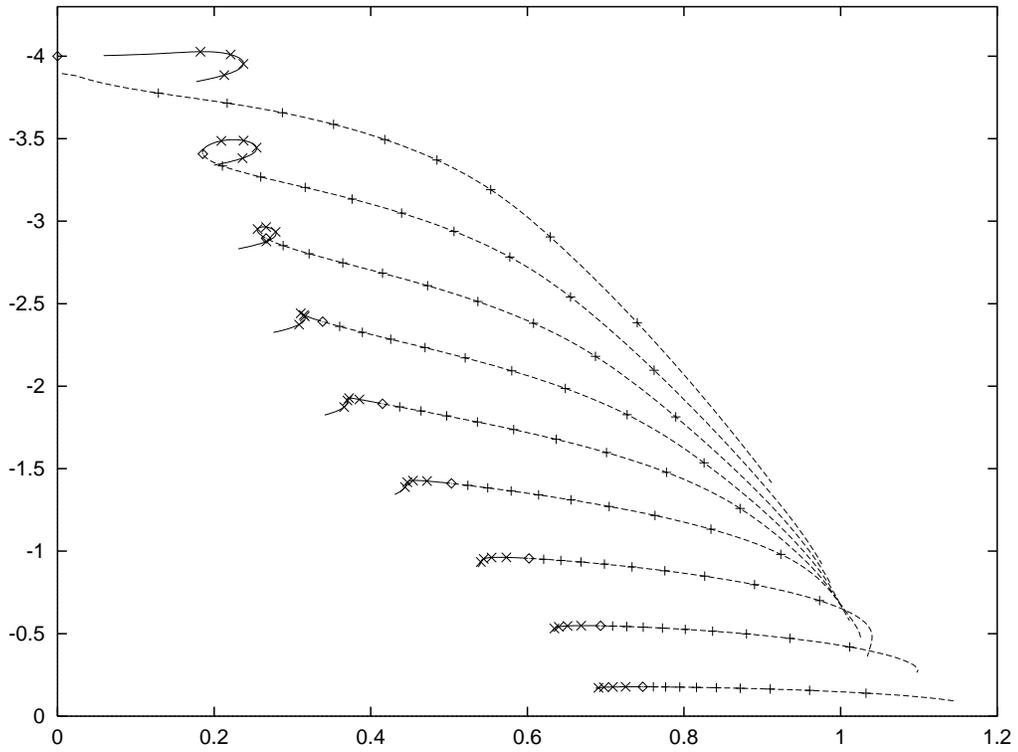}
\end{center}
\caption{
The first nine m=-1 and m=+1 gravitational quasinormal modes 
for l=2 are plotted.
The diamonds are the Schwarzschild quasinormal frequencies and the right
part starting from the Schwarzschild frequency is the branch of the
m=+1 modes, which is depicted by the dashed lines.
The left part is the branch of the m=-1, which is depicted by the solid lines.
We marked the values of the a=0.05, 0.1, 0.15, ...., 0.45 Kerr
frequencies for the m=+1
branch and a=0.1, 0.2, 0.3, 0.4 for the m=+1 branch.
It is noticed that the ninth mode comes out of the imaginary axis
that corresponds to the algebraically special mode of the Schwarzschild
black hole.
}
\end{figure}

\newpage

\begin{figure}
\begin{center}
  \leavevmode
  \epsfysize=10.0cm
  \epsfbox{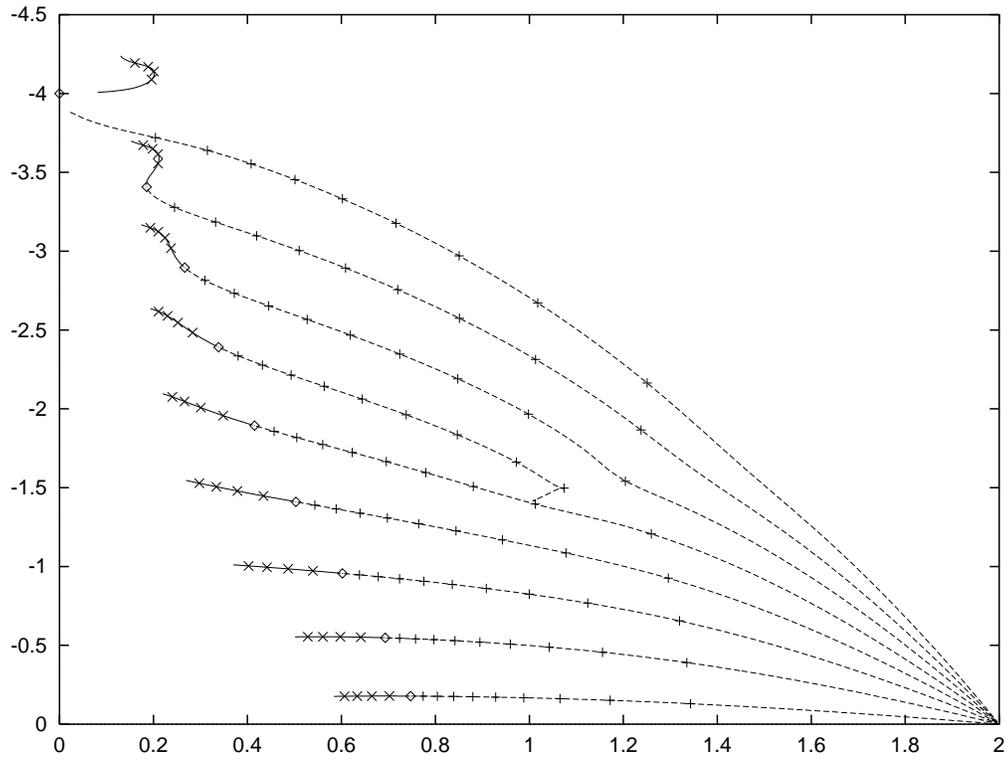}
\end{center}
\caption{
The first nine l=2 and m=-2,+2 gravitational quasinormal modes are plotted.
All the modes of the m=+2 branch
except the sixth overtone (n=5) converge into the
critical frequency on the real axis.
This accumulation of modes is in agreement with Detweiler's result.
But our numerical results suggest that some of the modes do not
cluster into the critical frequency.
}
\end{figure}

\newpage

\begin{figure}
\begin{center}
  \leavevmode
  \epsfysize=10.0cm
  \epsfbox{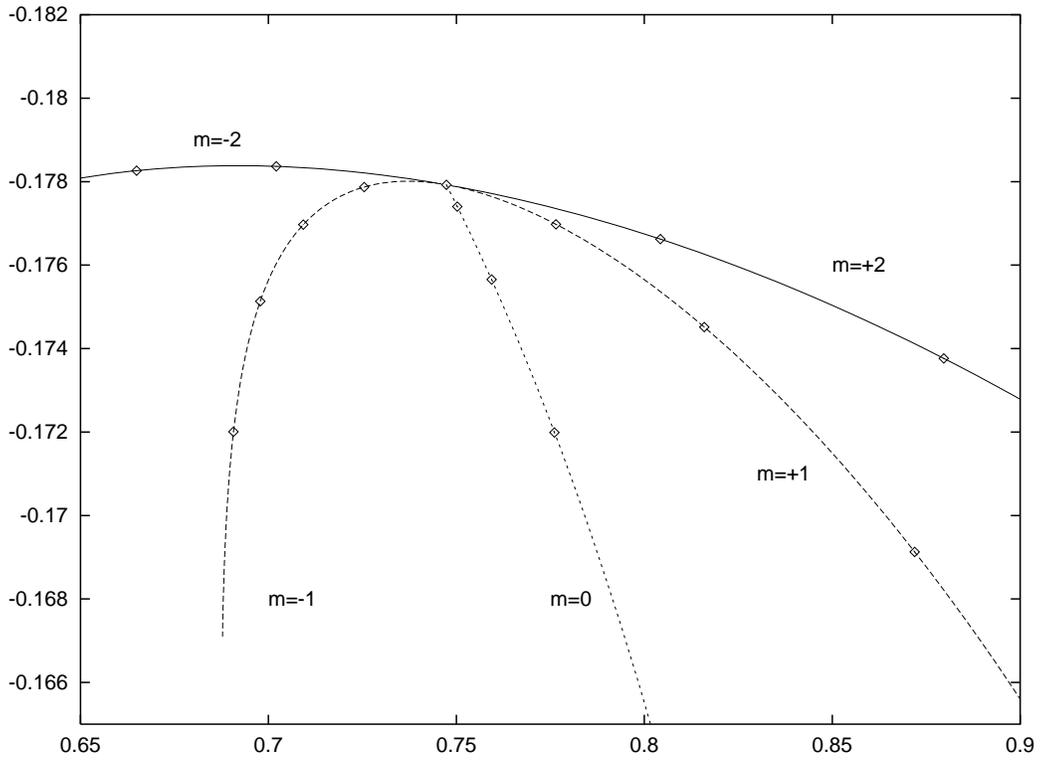}
\end{center}
\caption{
The l=2 gravitational fundamental quasinormal modes for any m are
zoomed up.
The branch of m=+1,-1 is tangent to the branch of m=+2,-2, but the
m=0 mode is not.
The marks correspond to the a=0 0.1, 0.2, 0.3, 0.4 values.
}
\end{figure}

\newpage

\begin{figure}
\begin{center}
  \leavevmode
  \epsfysize=10.0cm
  \epsfbox{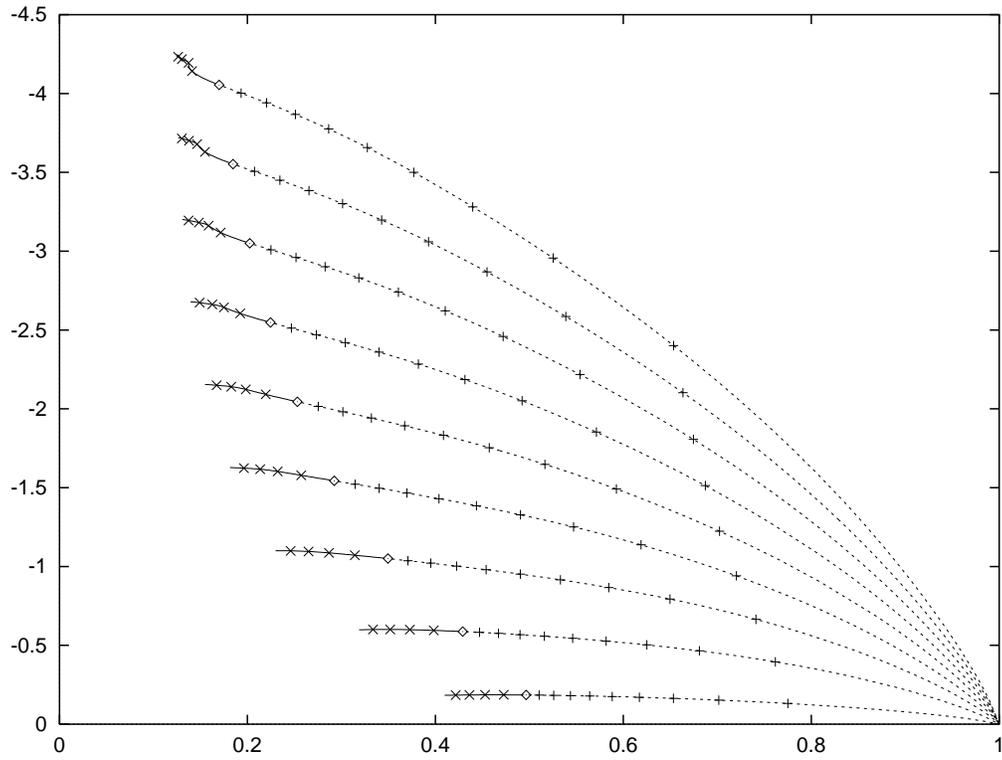}
\end{center}
\caption{
The first nine m=-1 and m=+1 quasinormal modes for the electromagnetic
waves in the Kerr geometry are shown.
This is very similar to the case of l=m=2 gravitational
quasinormal modes, however
all the mode up to n=8
converge into the critical frequency
in contrast to the gravitational frequencies.
}
\end{figure}

\newpage

\begin{figure}
\begin{center}
  \leavevmode
  \epsfysize=10.0cm
  \epsfbox{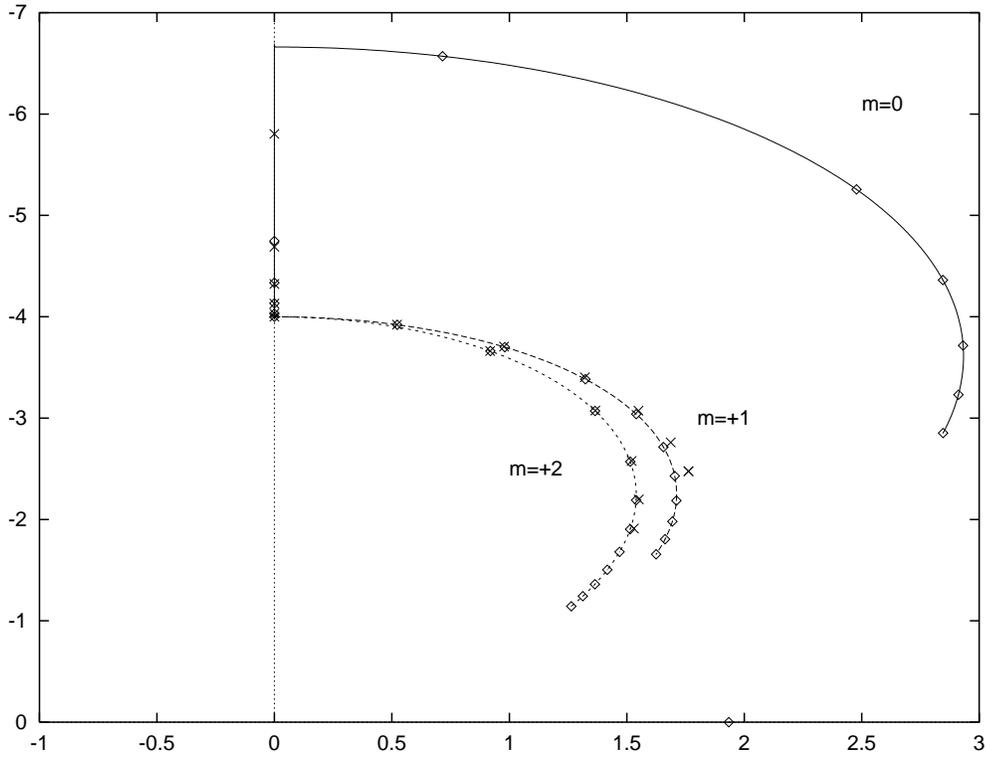}
\end{center}
\caption{
The algebraically special modes of m=0, +1 and +2 for l=2.
There also exist another branches of modes symmetric about the real axis.
The mode of m=-1 and m=-2 are in the left half plane, which are obtained
by reflecting the m=+1 and m=+2 modes about the imaginary axis, respectively.
The m=0 algebraically special mode first goes up vertically and then produces
two branches, one in the right half of the complex plane and the other
in the left half of the complex plane, which is not shown in this figure.
The corresponding Kerr parameter is a=0.247222 and the pure imaginary
frequency is about -6.681i.
}
\end{figure}

\newpage

\begin{table}
\begin{center}
\begin{tabular}{|c|c|c|c|}
\hline
 a    & m=0         & m=1               & m=2\\ \hline
 0.00 & (0, -4.0000)  & (0.0000, -4.0000)   & (0.0000, -4.0000)  \\ \hline
 0.05 & (0, -4.03107) & (0.522581,3.92142)  & (0.918857,-3.66121)\\
      & (0, -4.0310)  & (0.5218, -3.9218)   & (0.9184, -3.6620)  \\ \hline
 0.10 & (0, -4.13218) & (0.980889,-3.69993) & (1.36523,-3.07054) \\
      & (0, -4.1304)  & (0.9772, -3.7056)   & (1.3662, -3.0750)\\ \hline
 0.15 & (0, -4.33511) & (1.32376,-3.38382)  & (1.51483,-2.57005) \\
      & (0, -4.3240)  & (1.3216, -3.4026)   & (1.5204, -2.5774)\\ \hline
 0.20 & (0, -4.74699) & (1.54036,-3.03936)  & (1.53973,-2.19077) \\
      & (0, -4.6906)  & (1.5498, -3.0738)   & (1.5508, -2.1986)\\ \hline
 0.25 & (0.715878,-6.57057) & (1.65595,-2.71464) & (1.51441,-1.90331) \\
      & (0, -5.8038)  & (1.6868, -2.7600)   & (1.5304, -1.9104)\\ \hline
 0.3  & (2.47766, -5.25615) & (1.70404,-2.42938) & (1.46928,-1.68066) \\
      &               & (1.7626, -2.4766)   &                  \\ \hline
 0.35 & (2.84578, -4.36141) & (1.71091,-2.18596) & (1.41732,-1.50403) \\
      &               & (1.53041,-1.9104)   &                  \\  \hline
 0.40 & (2.93170, -3.71616) & (1.69372,-1.97997) & (1.36428,-1.36084) \\ \hline
 0.45 & (2.91142, -3.23046) & (1.66300,-1.80540) & (1.31270,-1.24256) \\ \hline
 0.50 & (2.84632, -2.85259) & (1.62505,-1.65663) & (1.26369,-1.14328) \\ \hline
\end{tabular}
\end{center}
\caption{The algebraically special modes calculated through the
continued fraction algorithm are compared with those obtained in
Chandrasekhar's paper.
The upper values are obtained through the continued fraction
method and the lower values are from his paper.
It seems his values are not accurate for large Kerr parameters,
owing to the numerical errors caused when getting the
separation constants.
}
\end{table}

\end{document}